\documentclass[11pt]{article}
\usepackage{hyperref}
\pdfoutput=1
\begin{document}
\title{Boundary layer evolution in turbulent Rayleigh-B\'{e}nard convection}
\author{Herwig Zilken$^1$, Mohammad S. Emran$^2$ and J\"org Schumacher$^2$ \\ \\
$^1$ J\"ulich Supercomputing Centre, Research Centre J\"ulich,\\D-55425 J\"ulich, Germany,\\
$^2$ Institute of Thermodynamics and Fluid Mechanics, \\Ilmenau University of Technology,
D-98684 Ilmenau, Germany}
\maketitle

\begin{abstract}
We present a fluid dynamics video which illustrates the dynamics of the velocity field in the boundary layer 
in a turbulent Rayleigh-B\'{e}nard convection flow. The data are obtained from direct numerical simulation.
\end{abstract}

\section{Video Description}
When a gas or fluid in a cylindrical cell of height $H$ and diameter $D$ is cooled from above and heated from below turbulent 
convection is triggered provided that the temperature difference between top and bottom plates is sufficiently
large. Crucial for the deeper understanding of the mechanisms of turbulent heat transport is the analysis of the physics in the
tiny boundary layers at the top and bottom \cite{Verzicco2012}. The video shows the dynamics of velocity field
in the thermal boundary layer. Data are obtained from direct numerical simulations of the Boussinesq equations (see 
\cite{Schumacher2012} for further details). The computational grid is given in cylindrical coordinates and contains 
$N_r\times  N_{\phi}\times N_z=513\times 1153\times 861$ grid points. The simulation parameters
are Prandtl number $Pr=0.7$, Rayleigh number $Ra=3\times 10^{10}$ and aspect ratio 
$\Gamma=D/H=1$. 

The movie shows at first the horizontal plane in which the streaklines are seeded. We display the temperature distribution and
observe a skeleton of hot sheet-like thermal plumes which turns from color into gray. 
This seeding plane is at about the thermal boundary layer thickness above the hot bottom plate. The streaklines are colored 
with respect to the velocity magnitude (blue=minimum, red=maximum). The evolution of the streaklines gives us an 
impression about the large-scale circulation which always builds up in a confined cell \cite{Chilla2012}. The second part
of the Video highlights the merging of two plume events close to the side wall. Our simulations show that the detachment of 
thermal plumes is accompanied by the formation of mini-tornados as they are shown here in the merging process. This zoom 
demonstrates clearly the locally fluctuating and three-dimensional  dynamics in the boundary layer. The implications for 
the boundary layer dynamics at even higher Rayleigh numbers and a transition to  turbulence need to be explored in the future.    

The computations have been conducted on 2048 cores of the Blue Gene/P JUGENE at the 
J\"ulich Supercomputing Centre under grant HIL02. We acknowledge support by the Deutsche 
Forschungsgemeinschaft within the Heisenberg Program and the Research Unit FOR 1182.

\end{document}